\documentclass[twocolumn ,amssymb,amsmath,nobibnotes,aps,nofootinbib]{revtex4}
\usepackage{amsmath}
\usepackage{bm}
\usepackage{graphicx}
\usepackage{slashed}
\usepackage{appendix}

\newcommand{\be}{\begin{equation}}
\newcommand{\ee}{\end{equation}}
\newcommand{\bea}{\begin{eqnarray}}
\newcommand{\eea}{\end{eqnarray}}
\newcommand{\bc}{\begin{center}}
\newcommand{\ec}{\end{center}}
\newcommand{\dd}{\ensuremath{\mathrm{d}}}

\begin{document}

\title{On the influence of a Chern-Simons term in the dynamical fermion masses in Reduced or Pseudo QED}

\author{Juan Angel Casimiro Olivares$^1$, Luis Albino$^2$, Ana Julia Mizher$^{2,3}$, Alfredo Raya$^{1,3}$}

\affiliation{$^1$ Instituto de F\'{\i}sica y Matem\'aticas, Universidad Michoacana de San Nicol\'as de Hidalgo. Edificio C-3, Ciudad Universitaria. Francisco J. M\'ijica s/n Col. Fel\'{\i}citas del R\'{\i}o. C.~P. 58040, Morelia, Michoac\'an, Mexico.\\
$^2$ Instituto de F\'{i}sica Te\'orica, Universidade Estadual Paulista, Rua Dr. Bento Teobaldo Ferraz, 271 - Bloco II, 01140-070 S\~ao Paulo, SP, Brazil.\\
$^3$Centro de Ciencias Exactas - Universidad del Bio-Bio.\\ Avda. Andr\'es Bello 720, Casilla 447, Chill\'an, Chile.
}

\begin{abstract}
Mixed dimensional theories have been used to describe condensed matter systems where fermions are constrained to a plane while the gauge fields they interact with remain four-dimensional. Here we investigate dynamical breaking of chiral symmetry in the framework of a mixed-dimensional theory that has been successful in describing graphene and other planar Dirac materials and has been dubbed as Pseudo or Reduced Quantum Electrodynamics. We explore the interplay between the gauge and fermion sector when a Chern-Simons term is considered and, by exploring the tensor structure of the fermion self energy,  we show that the radiative corrections may induce both, a Dirac and a Haldane mass term. Furthermore, by solving the corresponding Schwinger-Dyson equation in a suitable truncation, we explore non-perturbatively the dynamical generation of fermion masses for different values of the electromagnetic coupling and the Chern-Simons coefficient. We also show that for definite parameter values, the contributions from each chirality cancel out and the chiral symmetry may be restored. Possible implications of this result for physical systems, in particular on what concerns the chiral magnetic effect, are discussed. 

\end{abstract}
\maketitle

\section{Introduction}

Quantum Electrodynamics in two spatial and one temporal dimensions  (QED$_3$) has been intensively studied in the past mainly because it shares two important features with Quantum Chromodynamics (QCD), chiral symmetry breaking (CSB) and confinement~\cite{QED3-1,QED3-2}. This allows to interpret such a theory as a QCD toy model and has been widely explored in literature~\cite{Appelquist:1986fd,Pisarski:1984dj,Appelquist:1988sr,Nash:1989xx,Atkinson:1989fp,Burden:1990mg,Pennington:1990bx,Burden:1991uh,Gusynin:1995bb,Maris:1996zg,Gusynin:2000zb,Bashir:2001vi,Bashir:2002dz,Bashir:2002sp,Gusynin:2003ww,Fischer:2004nq,Bashir:2004yt,Bashir:2005wt,Bashir:2008fk,Bashir:2009fv,Bashir:2011vg,Raya:2013ina,Gusynin:2016som,Kotikov:2016wrb,Kotikov:2020slw}.
 Under laboratory conditions of temperature and pressure the chiral symmetry of QCD is spontaneously broken. This symmetry breaking dynamically generates the masses of the constituent-quarks and is ultimately responsible for the masses of light baryons and pseudoscalar mesons. However, under extreme conditions of the external parameters this symmetry may be restored, making the constituent-quark masses to vanish. To understand the nature and dynamics of this phase transition is one of the major goals of heavy ion physics, with deep implications on what we know on the early universe.

In this context, QED$_3$ has been used as a tool to explore CSB. Schwinger-Dyson equations (SDE) offer a suitable non-perturbative framework to capture the essence of this symmetry breaking, considering improved dressing of vertices and propagators. In a $1/N$ expansion for pure QED$_3$ it has been shown that in most cases there is a critical number of fermion flavors, above which the chiral symmetry breaking is restored \cite{Appelquist:1988sr,Appelquist:1986fd,Nash:1989xx,Hands:2004bh}. Beyond this approximation, refined vertices and vacuum polarization effects agree with this findings in general, refining the details of the critical line~\cite{Fischer:2004nq}. These results point out that what actually defines the existence of this criticality is the infrared behavior of the fermion wave function renormalization~\cite{Bashir:2008fk}.

Physics on a plane is interesting on its own. It permits the inclusion of a Chern-Simons (CS) term in the gauge sector which induces fractional (anyon) statistic and parity (and time-reversal) non-invariance, among other interesting effects (see, for instance, Refs.~\cite{khare,lerda,wilczek} and references therein).  Regarding CSB in QED$_3$, the impact of adding the CS term in the photon sector was done by using different truncations of the SDE either in an appropriate non-local gauge \cite{Kondo:1994bt,Kondo:1994cz} or in the Landau gauge \cite{Hofmann:2010zy}. These results coincide that, besides a critical number of families, there is also a critical value for the Chern-Simons parameter $\theta$ which induces chiral symmetry restoration. A critical line as a function of both parameters can be defined and the literature~\cite{Kondo:1994bt,Kondo:1994cz,Hofmann:2010zy} points towards a first order transition.

With the experimental realization of graphene in 2004, QED$_3$ met a more direct application. Besides being constrained to a plane, the charge carriers in this material in the low-energy regime obey a linear dispersion relation and carry a chirality degree of freedom. Because of that, QED$_3$ was identified as the continuous limit of the tight-bind description of monolayer graphene~\cite{Gusynin:20x}. This relativistic-like character of the quasi-particles in graphene - which subsequently was found in a few other Dirac/Weyl materials~(see, for instance, \cite{2Dmat} and references therein) - has opened for the possibility of testing in a table top environment certain phenomena that were initially predicted in the context of high energy physics, like the Klein paradox \cite{klein} and the Zitterbewegung \cite{Zitterbewegung}. 

Although QED$_3$ has been shown to correctly describe the charge carriers in Dirac materials, interaction between these electrons and external electromagnetic fields cannot be correctly described by the full-fledged dimensionally reduced theory including gauge fields in $(2+1)$ dimensions. This is due to the fact that although the fermionic quasi-particles are constrained to the plane, the gauge fields live in a bulk of higher dimensions and it is necessary to consider a mixed-dimensional theory to correctly describe this interaction. A procedure to deal with this condition was developed considering a four-dimensional gauge field interacting with fermion fields living in the boundary sheet $x_3=0$. This method first appeared in \cite{Marino:1992xi} as pseudo-QED (PQED) and later was developed in~\cite{Gorbar:2001qt}, being dubbed reduced QED (RQED). We adopt the notation P(R)QED.

Here, we investigate chiral symmetry breaking in the framework of P(R)QED. This was done previously in the literature \cite{Gorbar:2001qt,Alves:2013bna,Nascimento:2015ola}, but in the present work we focus on the influence of the Chern-Simons term on the dynamical mass generation in P(R)QED and its consequences to chiral symmetry. Pioneering investigation on CSB critical coupling in QED$_4$ was done in \cite{Fukuda:1976zb}, where the author found that CSB occurs above a critical coupling $\alpha_c=\pi/3$. In \cite{Miransky:1986ib,Fomin:1984tv}, the author proposed that the dynamically generated mass should obey a scaling law $\propto \Lambda e^{-\pi/\sqrt{\alpha/\alpha_c-1}}$, what is known as Miransky scaling. In \cite{Alves:2013bna,Nascimento:2015ola,Cristobal} the authors perform a similar calculation for pure P(R)QED, obtaining a critical value $\alpha_c=\pi/8$. Thermal effects were further considered in~\cite{Nascimento:2015ola,Cristobal} and it was explicitly shown how $\alpha_c$ depends on the temperature.

Our goal in the present paper is to explore the parameter space and detect for which values of the coupling and the CS-$\theta$ parameter CSB occurs. The manuscript is organized as follows: in section \ref{lagrangian} we present the Lagrangian and discuss the Feynmann rules. 
In section \ref{SD} we solve the corresponding Schwinger-Dyson equation under the rainbow-ladder approximation in Landau gauge and neglecting wavefunction renormalization effects. We observe that, for suitable values of the parameters, two types of mass terms emerge. Finally, in section \ref{conclusions} we draw our conclusions and comment possible implications of our results on high-energy related condensed matter phenomena.

\section{Lagrangian and Feynman rules}
\label{lagrangian}

P(R)QED can be defined through the action:
\begin{eqnarray}\label{qed}
  \mathcal{S}_{\text{QED}_4} &=& \int \dd^4x \left[-\frac{1}{4}F_{\mu\nu}F^{\mu\nu}+ej_\mu A^\mu\right]+S_{gf}\,,
\end{eqnarray}
which describes a four-dimensional gauge field interacting with a three-dimensional fermionic conserved current, given by 
\begin{eqnarray}\label{current}
  j^{\mu} &=& \left\{\begin{array}{cc}
              i\bar \psi \gamma^\mu \psi \delta(x_3) &  ~\text{for}~\mu=0,1,2\,, \\
              0& ~\text{for}~\mu=3\,.
            \end{array}\right.
\end{eqnarray}
Fermion dynamics can be included by adding up a three-dimensional kinetic term  $\int \dd^3x \bar\psi i \slashed{\partial}\psi$. 

The procedure consists on integrating out the four-dimensional gauge field followed by an integration over the third space direction, perpendicular to the sheet where the fermion dynamics evolves. After that, one re-introduces the gauge fields, where the new field is consistently defined in three dimensions. This procedure leads to the following effective action: \begin{eqnarray}\label{rqed}
\mathcal{S}_{\text{RQED}_3}=\int \dd^3 x \left[ \frac{1}{2}   F_{\mu\nu} \frac{1}{\sqrt{-\Box^2}} F_{\mu\nu}+\bar{\psi} i\slashed{D}\psi  \right]+S_{gf},
\end{eqnarray}
where $F_{\mu\nu}$ is the electromagnetic field tensor, $A^\mu$ contained in the covariant derivative is the (three-dimensional) gauge field and $\psi$ the (four-component) fermion field. $S_{gf}$ is a gauge fixing term. Although P(R)QED is notably non-local, it has been shown that it respects causality and its Green functions are well defined \cite{doAmaral:1992td}.

Working with a linear gauge fixing, we can deduce propagators and vertices from the following Lagrangian:
	\begin{equation}
		\mathcal{L}=-\frac{1}{4}F^{\mu\nu}\frac{2}{(-\Box)^{1/2}}F_{\mu\nu}+\bar{\psi}(i\gamma^\mu\partial_\mu+e\gamma^\mu A_\mu)\psi+\frac{1}{2\zeta}(\partial\cdot A)^2,
		\label{ec}
	\end{equation}
	where indices run from $0$ to $2$, $e$ is the electric charge and $(-\Box)$ is the D'Alambertian operator, which appears in the Lagrangian with fractional power and the last term is the gauge fixing term with $\zeta$ de gauge fixing parameter.
	
The Euclidean Clifford algebra associated to the gamma matrices, $\{\gamma_\mu,\gamma_\nu\}=2\delta_{\mu\nu}$, has two possible irreducible two-dimensional representations. A parity transformation swaps massive fermions from one representation into the other so that none of them are parity invariant. Due to the fact that in odd dimensions it is not possible to define a $\gamma_5$ matrix - one that is constructed from the anticommuting matrices in a given representation and anti-commutes with all of them - it is not possible to define chirality. It is possible, however, to merge the particles in different representations and define a new, reducible, four-component representation \cite{deJesusAnguianoGalicia:2005ta}. This is actually the usual procedure to define chirality in graphene \cite{Gusynin:20x}, where the degrees of freedom of sublattices and valleys are represented in a single four-component structure and chirality can then be defined associated to the later. 

In what follows we consider the chiral limit, which amounts to set all fermion masses to zero. 
Furthermore, when solving the Schwinger-Dyson equation for the fermion propagator, we find it convenient to work with the four-component reducible representation of the fermion fields, as explained below.



	The photon propagator derived from (\ref{ec}) is:
	\begin{equation}
		\Delta^0_{\mu\nu}(q)=\frac{1}{2q}\Big(\delta_{\mu\nu}-\frac{q_\mu q_\nu}{q^2}\Big)+\frac{\zeta}{q^2}\frac{q_\mu q_\nu}{q^2}.
		\label{RQED_photonprop}
	\end{equation}
	One can notice that it presents a softer infrared behavior than the photon propagator in QED in four dimensions.
	%
	The massless fermion propagator is  given by:
	\begin{equation}
	S_0^{-1}(p)=-\gamma^\mu p_\mu.
	\end{equation}

The focus of this work is to explore the consequences of coupling P(R)QED to a Chern-Simons term. Chern-Simons (CS) belong to a class of gauge theories that can be only defined in odd dimensions. Besides being a breakthrough in mathematical physics, they have been useful to describe several condensed matter phenomena, from the fractional quantum Hall effect~(~see \cite{khare,lerda} and references within), to the investigation on high $T_c$ superconductivity~(see \cite{wilczek} and references within). The CS Lagrangian is constructed as a gauge theory, alternative to the Maxwell Lagrangian, when one imposes Lorentz and gauge invariance, as far as locality, in odd dimensions:

%
\begin{equation}
\mathcal{L}_{CS}=\frac{\kappa}{2}\varepsilon^{\mu\nu\rho}A_\mu \partial_\nu A_\rho - A_\mu J^\mu.
\label{CSLagrangian}
\end{equation}
Note that under a gauge transformation the CS-Lagrangian remain unchanged apart from a total space-time derivative. Therefore, the corresponding action $S_{CS}=\int d^3 x\  \mathcal{L}_{CS}$ is gauge invariant as far as we are allowed to neglect boundary terms.
	
In order to explore the influence of a CS term in the chiral symmetry we add it up to the Lagrangian (\ref{ec}), yielding: 
\begin{eqnarray} \nonumber
		\mathcal{L}_{RQED}^ {CS}&=&-\frac{1}{4}F^{\mu\nu}\frac{2}{(-\Box)^{1/2}}F_{\mu\nu}+\bar{\psi}(i\gamma^\mu\partial_\mu+e\gamma^\mu A_\mu)\psi\\
		&+&\frac{1}{2\zeta}(\partial\cdot A)^2 + \frac{-i\theta}{4}\varepsilon^{\mu\nu\rho}A_\mu \partial_\nu A_\rho,
		\label{final_Lagrangian}
	\end{eqnarray}
	
	The bare photon propagator derived from the above Lagrangian can be written as \cite{Dudal:2018mms}:
\begin{eqnarray}\label{photon_prop_tree}\nonumber
\hat \Delta_{\mu\nu}(\vec{q})&=&\frac{1}{2q}\frac{1}{(1+\theta^2)}\left(\delta_{\mu\nu}
-\frac{q_\mu q_\nu}{q^2}\right)\\&&+\frac{\zeta}{q^2}\frac{q_\mu q_\nu}{q^2}-\frac{1}{2q^2}\frac{\theta}{(1+\theta^2)}\epsilon_{\mu\nu\rho}q^{\rho}
 \;.
 \label{photon_prop}
\end{eqnarray}
We can verify from the above expression that the $\theta$ parameter is dimensionless so, in contrast to QED$_3$, it cannot be associated to the so-called topological photon mass \footnote{Actually, in the Abelian case there is no real topology
involved and the term “topological” is used for historical reasons based on its non-Abelian counterpart.}. This can be inferred from the fact that P(R)QED posses the remarkable property of scale invariance \cite{Dudal:2018pta}. Comparing to the propagator of pure P(R)QED (\ref{RQED_photonprop}), the presence of a CS term induces a T-odd contribution in the propagator, besides a normalization of the propagator. This is explicitly reflected in the mass generation we  present below in section \ref{SD}.

\section{Non-perturbative solution}
\label{SD}

In order to explore the dynamical  generation of fermion mass in R(P)QED, we work within the Schwinger-Dyson equations (SDE) framework, which has vastly been made use of for the issue of CSB in QCD (see \cite{CraigRoberts:2020,AAguilar,FischerRev,Horn} for recent reviews). Such a framework consists of an infinite tower of coupled equations among the $n$-point Green functions of the theory. The corresponding equations for the two-point functions are
	\begin{eqnarray} 
		S^{-1}(p)&=&S_0^{-1}(p)-\Xi(p),\\
		\Delta^{-1}_{\mu\nu}(p)&=& \Delta^{-1}_{0\mu\nu}(p)-\Pi_{\mu\nu}(p)
		\label{ec1}
	\end{eqnarray}
corresponding to the fermion and photon propagators respectively. The index zero stands for the bare propagator and $\Xi(p)$ and $\Pi_{\mu\nu}(p)$ are the fermion and photon self-energies.  In what follows we work in the quenched limit - neglecting fermion-loop contribution for the vacuum polarization. Besides that,  we use the rainbow-ladder truncation - bare fermion-photon vertex - to decouple the SD equations. This is a common truncation adopted in the literature to capture the traits of CSB in different quantum field theories. A more refined truncation schemes demands satisfaction of the underlying symmetries of R(P)QED.
The general solution to the SDE for the fermion propagator in  (\ref{ec1}) is
	\begin{equation}
		S^{-1}(p)=-A(p)\gamma^\mu p_\mu+\Sigma(p).
	\end{equation}
We adopt the rainbow-ladder truncation by selecting the tree-level fermion-photon vertex and photon propagator. Furthermore, we neglect the wavefunction renormalization setting $A(p)=1$.
In this fashion, we have the following form for the mass function
	\begin{equation}
		\Sigma(p)=4\pi\alpha\int \frac{d^3 k}{(2\pi)^3}\frac{\Sigma(k)}{k^2+\Sigma^2(k)}\frac{1}{q},
		\label{ec3}
	\end{equation}
	where $\alpha=e^2/4\pi$ is the coupling. Notice that this equation has the same functional form as Eq.~(22) of Ref~\cite{Gorbar:2001qt}. However, the said equation has been derived on very different grounds. It was obtained by including vacuum polarization effects at the leading order of the $1/N_f$ approximation and working in a non-local gauge. As such, the coupling
	\begin{equation}
	   \alpha \to \lambda = \frac{4\alpha}{3\left( 1+\frac{\pi\alpha N_f}{4}\right)}
	\end{equation}
can serve now to identify the critical number of fermion families $N_f$ for chiral symmetry restoration.

When including the CS term, it is interesting to work with projected right and left handed fermions as $\psi_\pm=\chi_\pm \psi$. Given $\tau=[\gamma_3,\gamma_5]/2$, the chiral projector is defined as $\chi_\pm=(1\pm\tau)/2$, which implies in the following properties: $\chi_\pm^2=\chi_\pm$, $\chi_+ \chi_-=0$ and $\chi_+ + \chi_-=1$. The advantage of this procedure is the following: in order to explore the non-perturbative mass generation  we allow for the general case where an ordinary Dirac mass $m_e\bar{\psi}\psi$ can emerge. On top that, there is another mass term that potentially may be generated and is particularly interesting to us, known as Haldane mass $m_o\bar{\psi}\tau\psi$, with $\tau=[\gamma_3,\gamma_5]/2$. This term does not break chiral symmetry, but rather violates parity (and time-reversal) and because the CS is parity violating we expect a relation between them. On the other hand, the Dirac mass behaves in the opposite way, preserving $\mathcal{P}$ and $\mathcal {T}$ symmetry, but, of course, breaks chiral symmetry. With these considerations,  the most general Lagrangian we consider for our analysis  is
\begin{eqnarray} \nonumber
		\mathcal{L}_{RQED}^ {CS}&=&-\frac{1}{4}F^{\mu\nu}\frac{2}{(-\Box)^{1/2}}F_{\mu\nu}+\bar{\psi}(i\gamma^\mu\partial_\mu+e\gamma^\mu A_\mu\\ \nonumber
		&+& m_e + \tau m_o)\psi
		+\frac{1}{2\zeta}(\partial\cdot A)^2 + \frac{-i\theta}{4}\varepsilon^{\mu\nu\rho}A_\mu \partial_\nu A_\rho,
		\label{massive_Lagrangian}
	\end{eqnarray}
from which the fermion propagator has the structure:
\begin{equation}
    S_F^{-1}(p)=A_e(p)\slashed{p} + A_o(p)\tau \slashed{p} - B_e(p) -B_o(p) \tau.
\end{equation}

These masses however do not correspond to poles in the propagator. The operators $\chi_\pm$ project the upper and lower two-component spinors, making explicit the poles of each fermion species. Acting with the chiral projectors on the fermion sector,  we have
\begin{equation}
    \mathcal{L}_F = \bar\psi_+ (i\slashed\partial - M_+)\psi_+ + \bar\psi_- (i\slashed\partial - M_-)\psi_-,
\end{equation}
where $M_+=m_e+m_o$ and $M_-=m_e-m_o$.

Given that, we start from the gap equation 
	\begin{equation}
		S^{-1}_\pm(p)=S^{-1}_{0\pm}(p)+4\pi\alpha\int\frac{d^3k}{(2\pi)^3}\gamma^\mu S_{\pm}(k)\gamma^\nu \Delta_{\mu\nu}(q),
		\label{projected_prop}
	\end{equation}
where
	\begin{eqnarray}
		S^{-1}_\pm(p)&=&(\slashed{p}+M_\pm(p))\chi_{\pm}\nonumber\\
		S_0^{-1}(p)&=&\slashed{p}\chi_{\pm}\nonumber\\
		S_{\pm}(k)&=&-\frac{\slashed{k}+M_{\pm}(k)}{k^2+M^2_{\pm}(k)}\chi_{\pm}
			\end{eqnarray}
and $\Delta_{\mu\nu}$ is the photon propagator given by Eq.(\ref{photon_prop}). Replacing these equation in Eq.(\ref{projected_prop}) and taking the traces, we reach the following gap equation
			\begin{eqnarray} \nonumber
				M_\pm(p)&=&2\pi\alpha\int\frac{d^3k}{(2\pi)^3}\Big[\frac{2M_{\pm}(k)}{k^2+M^2_{\pm}(k)}\frac{1}{q(1+\theta^2)}\\
				&\mp&\frac{1}{q^2}\frac{\theta}{1+\theta^2}\frac{k\cdot q}{k^2+M^2_{\pm}(k)}\Big].
				\label{mpm}
			\end{eqnarray}
Performing a straightforward and standard calculation (detailed in  the appendix \ref{appendixA}), the gap equation evolves to the following differential equation:
			 \begin{eqnarray} \nonumber
			 	p^2M_\pm''(p)+2pM_\pm'(p)+\frac{2\alpha}{\pi(1+\theta^2)}M_\pm(p)\\
			 	=\mp\frac{2\alpha\theta}{\pi(1+\theta^2)}\Big[\frac{5p}{3}-\frac{6p^2}{\kappa}+\frac{\Lambda^3}{9p^2}-\frac{p}{9}\Big],
			 	\label{diff_eq}
			 \end{eqnarray}
with the boundary conditions
            \begin{equation}
             p^2M'(p)\Bigg|_{p\to 0} =0, \qquad (p M'(p)+M(p))\Bigg|_{p=\Lambda}  =0,
            \end{equation}
and whose solution is given by:
			 \begin{eqnarray}
			 	M_\pm(p)&=&p^{-\frac{1}{2}\sqrt{1-\frac{\alpha}{\alpha_c}}-\frac{1}{2}} \left(c_2 p^{\sqrt{1-\frac{\alpha}{\alpha_c}}}+c_1\right)\nonumber\\
			 	&+&f(\theta, p)\nonumber\\
			 	&=& \Lambda e^{-\pi/\sqrt{\alpha/\alpha_c-1}}+f(\theta,p),
			 	\label{SD_solution}
			 	 \end{eqnarray}
where $\alpha_c=\pi(1+\theta^2)/8$ and
\begin{eqnarray} \nonumber
f(\theta, p)&=&\Bigg[\pi  \left(\theta ^2+1\right) \left(\mp \frac{2\alpha  \theta }{\theta ^2+1}\right) \bigg(\kappa \left(\alpha +3 \pi  \left(\theta ^2+1\right)\right)\\ \nonumber
&&\left(\Lambda ^3+14 p^3\right)-54 p^4 \left(\alpha + \pi  \left(\theta ^2+1\right)\right)\bigg)\Bigg]\\
&\times&\frac{1}{18 \kappa p^2 \left(\alpha + \pi  \left(\theta ^2+1\right)\right) \left(\alpha +3 \pi  \left(\theta ^2+1\right)\right)}.
\label{f}
\end{eqnarray}

\begin{figure}
\includegraphics[width=8cm]{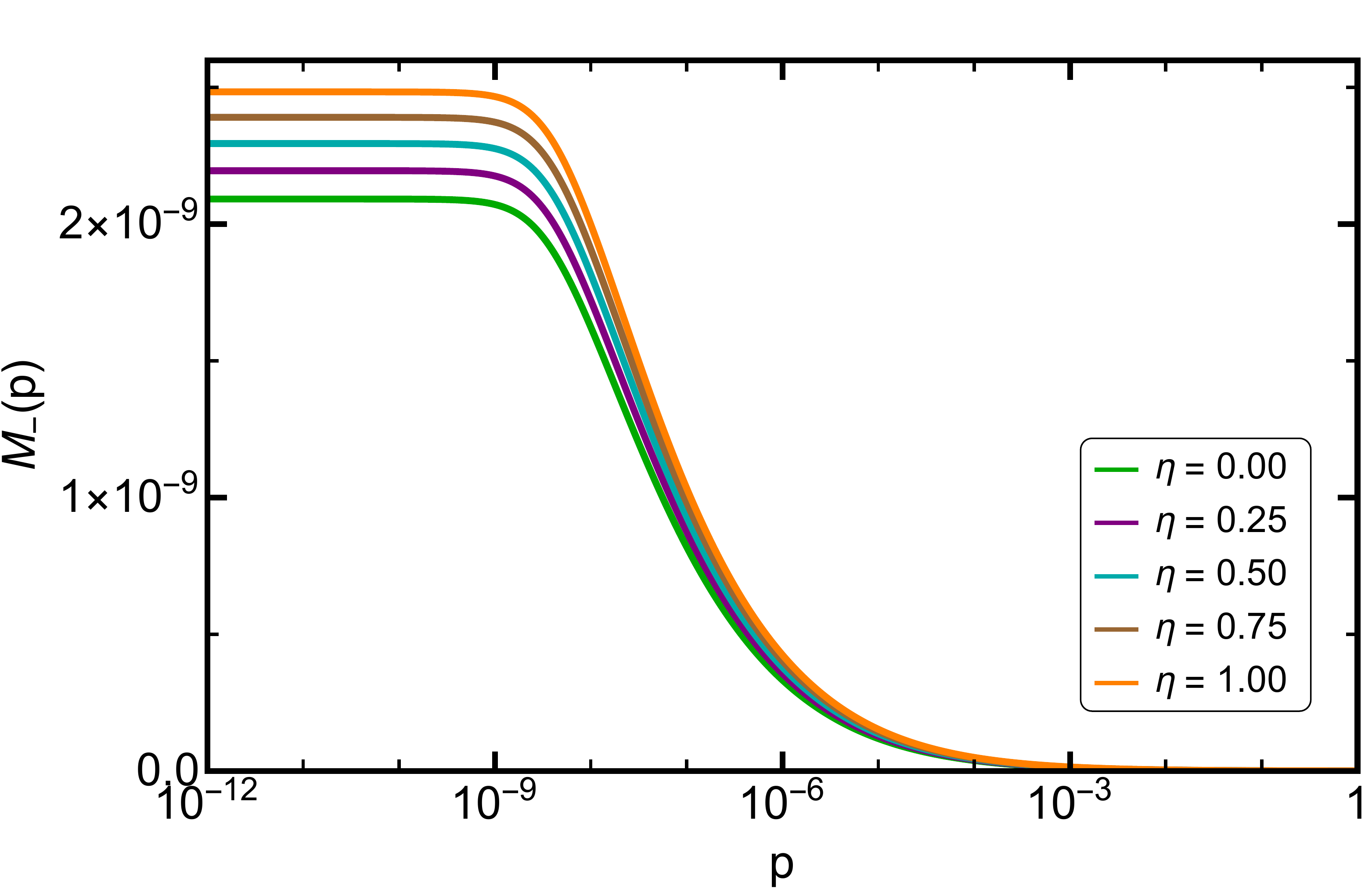}
\caption{Mass function $M_-$ plotted as a function of momentum. The different curves correspond to different values of the CS parameter $\theta=\eta \theta_c$ with $0\leq\eta\leq1$ and fixed value of the coupling $\alpha=1.07 \alpha_c$.}
\label{fig:Mminus_p}
\end{figure}

\begin{figure}
\includegraphics[width=8cm]{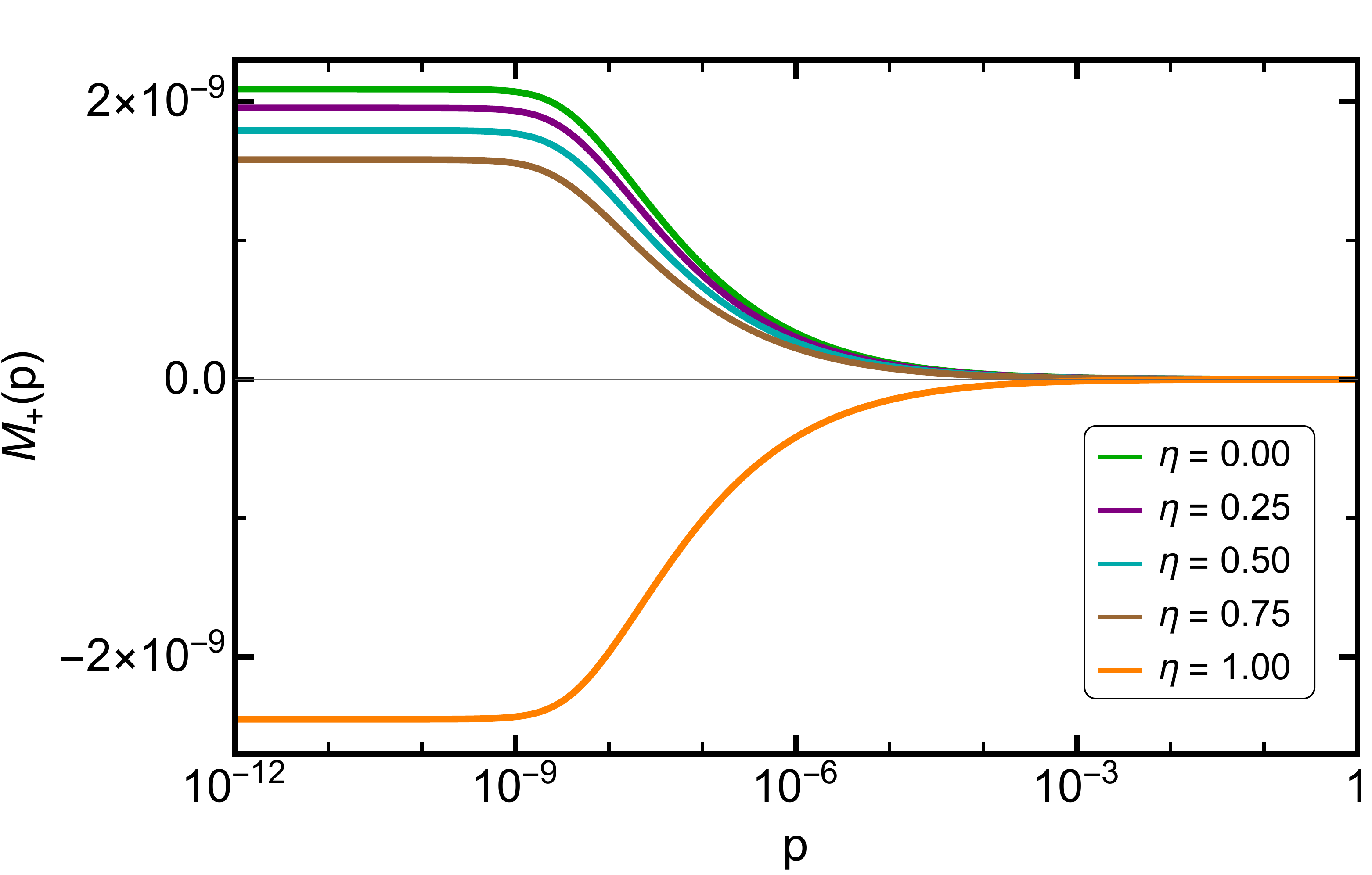}
\caption{Mass function $M_+$ plotted as a function of momentum. The different curves correspond to different values of the CS parameter $\theta=\eta \theta_c$ with $0\leq\eta\leq1$ and fixed value of the coupling $\alpha=1.07 \alpha_c$.}
\label{fig:Mplus_p}
\end{figure}

\begin{figure}
\includegraphics[width=8cm]{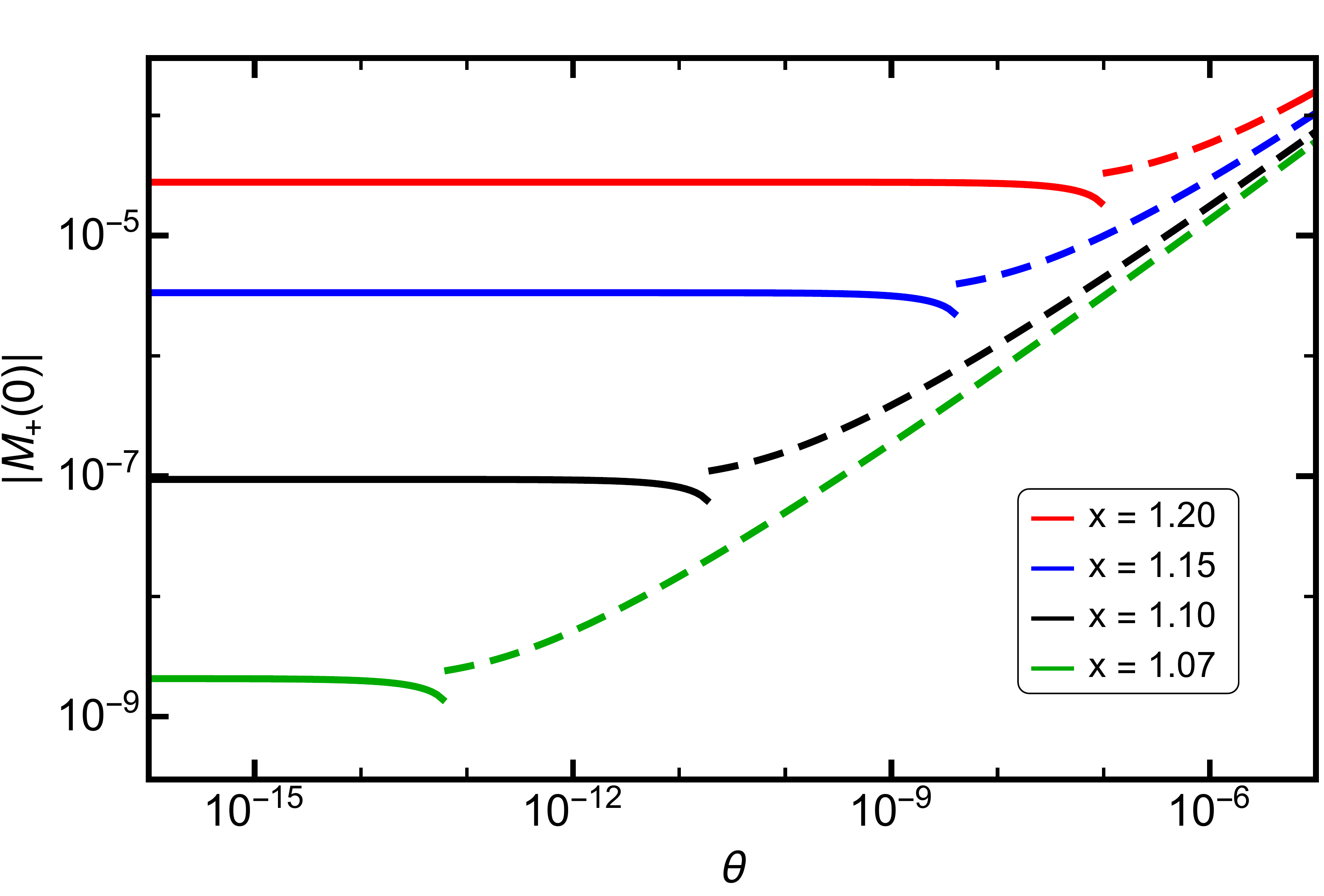}
\caption{The absolute value of the pole of the right fermion propagator $M_+$ plotted as a function of the CS parameter $\theta$. Above $\theta_c$ the mass $M_+$ becomes negative and here we flipped its sign for a better visualization. The different curves correspond to different values of the coupling $\alpha=x \alpha_c$.}
\label{fig:m_plus}
\end{figure}

\begin{figure}
\includegraphics[width=8cm]{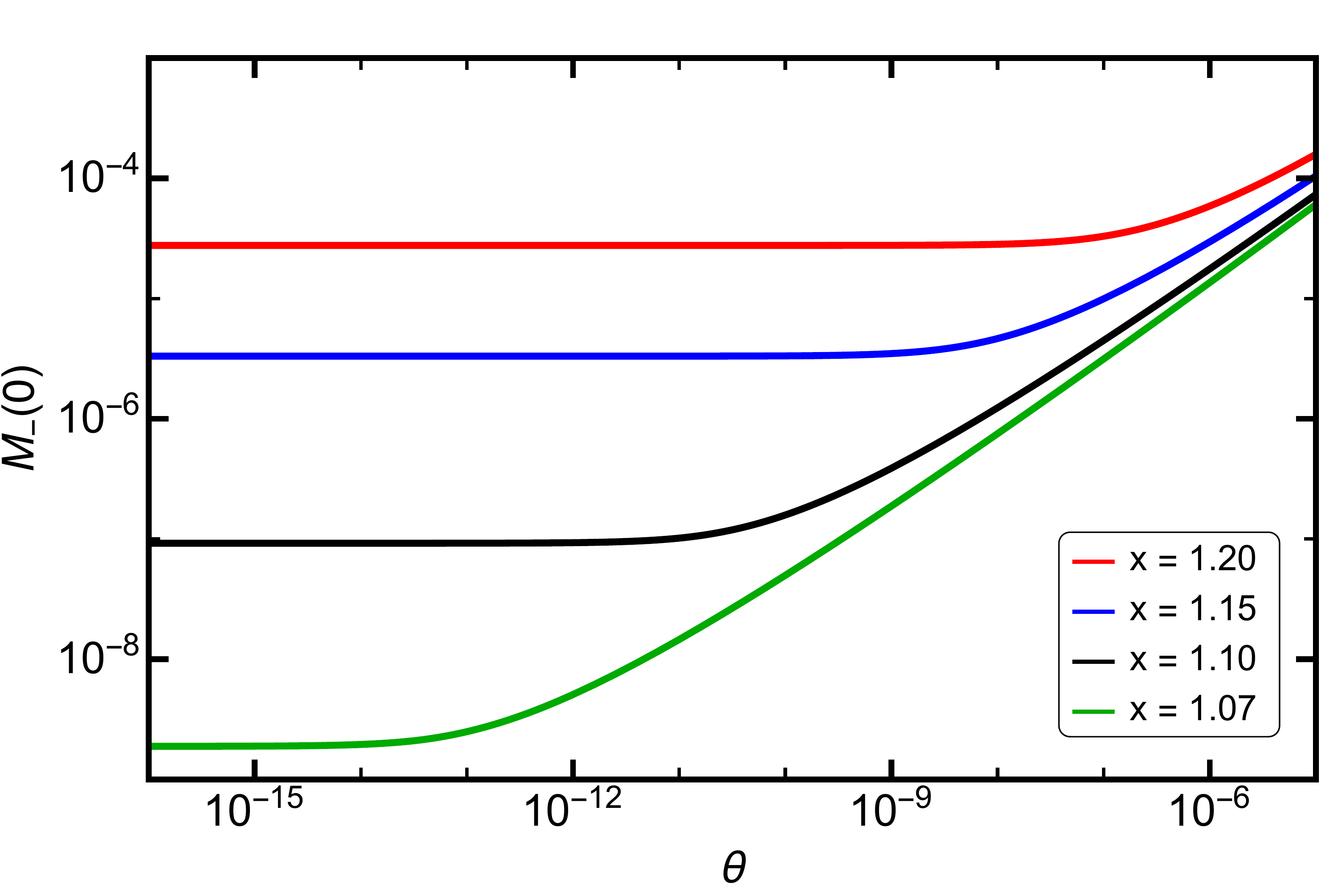}
\caption{The pole of the left fermion propagator $M_-$ plotted as a function of the CS parameter $\theta$. The different curves correspond to different values of the coupling $\alpha=x \alpha_c$.}
\label{fig:m_minus}
\end{figure}

\begin{figure}
\includegraphics[width=8cm]{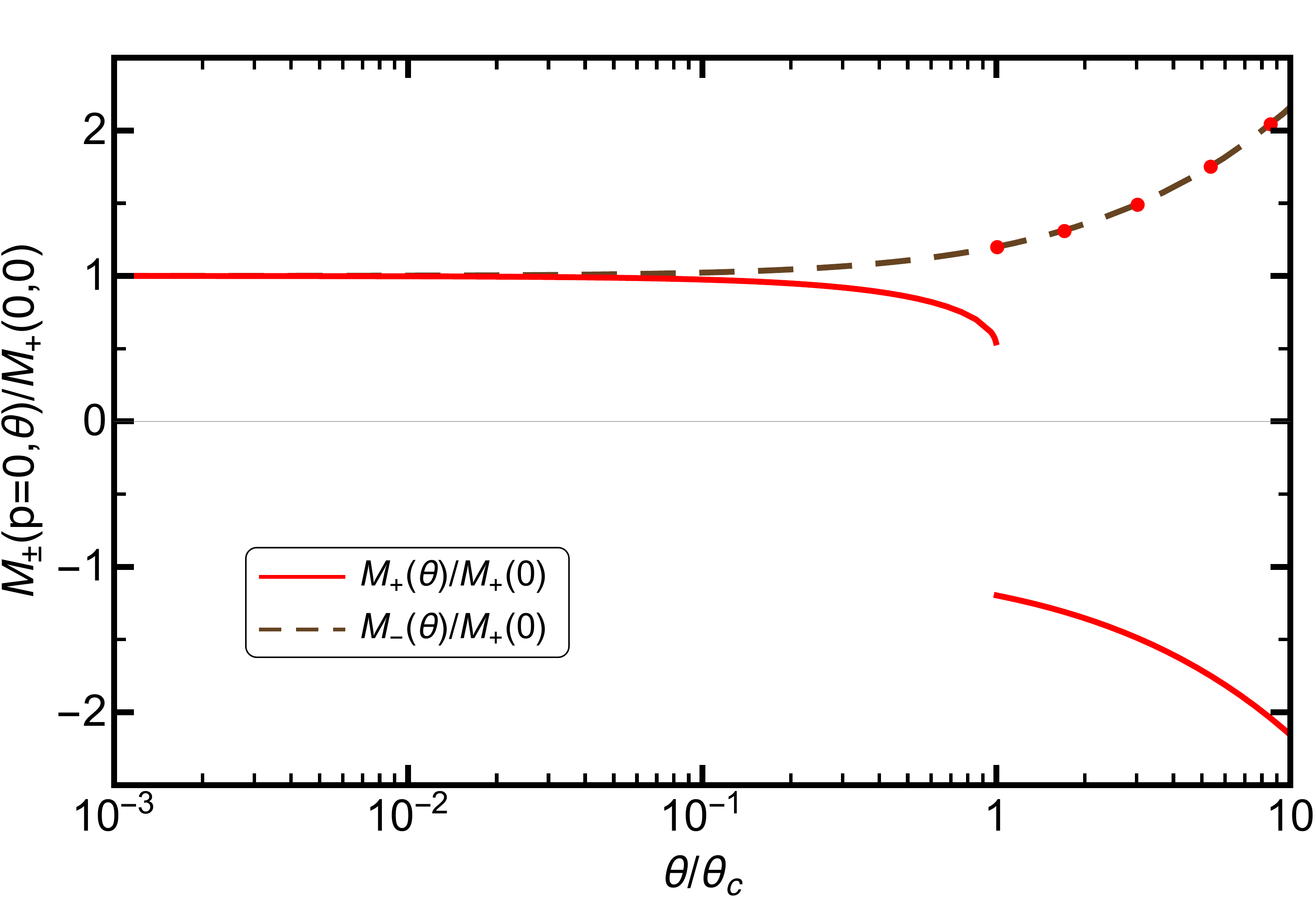}
\caption{$M_\pm$ re-scaled by their $\theta=0$ value. The full line corresponds to $M_+$ and dashed line $M_-$. The red dots are the absolute value of $M_+$ above the critical $\theta$.}
\label{fig:scaled_mass}
\end{figure}

We can identify the Miransky scaling law in the first term of the second row of Eq.(\ref{SD_solution}). This term corresponds to the solution of ordinary R(P)QED obtained from a homogeneous version of the differential equation (\ref{diff_eq}). The only difference from the first piece in the present solution to the solution of pure R(P)QED is a re-scaling of the critical coupling as $\alpha_c(1+\theta^2)$, whose limit $\theta\rightarrow 0$ brings it straightly to the usual critical coupling. Besides that, the inhomogeneity in the right hand side of Eq.(\ref{diff_eq}), brought by the extra CS piece in the photon propagator Eq.(\ref{photon_prop}), is responsible for the emergence of $f(\theta,p)$.

In Figs. \ref{fig:Mminus_p} and \ref{fig:Mplus_p} we plot $M_-$ and $M_+$ as a function of the external momentum $p$, respectively, for several values of $\theta=\eta \theta_c$, with $0\leq\eta\leq1$ for fixed $\alpha=1.07 \alpha_c$, which corresponds to $\theta_c=6.6\times 10^{-14}$. We notice that for small momentum both masses are finite and approximately $p$-independent. For larger momentum, the height of the mass functions start to decrease. This remains true for different values of $\theta$ as far as the coupling considered is above its critical value. For couplings below $\alpha_c$, the solution oscillates very rapidly around zero, indicating that the masses are vanishing. Therefore, Figs. \ref{fig:Mminus_p} and \ref{fig:Mplus_p} indicate that masses can be dynamically generated in CS-R(P)QED, and their magnitude is $\theta$ dependent. This dependence happens in opposite ways for $M_+$ and $M_-$: while the value of $M_-$ is incremented as $\theta$ increases, $M_+$ decreases, reaching a critical point at $\theta_c$ when it changes sign abruptly, signaling a first order phase transition. In order to examine the consequences on the chiral and parity symmetries we must check how this behavior reflects in $m_e$ and $m_o$.

The height of the mass functions in the deep infrared, $M_\pm(0)$, which measure the amount of dynamically generated mass are depicted in Figs. \ref{fig:m_plus} and \ref{fig:m_minus} for a fixed value of $\alpha=x\alpha_c$. Above $\theta_c$, the value of $M_+(0)$ undergoes a discontinuity and jumps assuming negative values. In the Fig.\ref{fig:m_plus} we have flipped the sign for $M_+$ above the discontinuity to lead the eye. In these figures we can clearly see the opposite behavior of the poles of right- and left-handed fermions as $\theta$ increases and appreciate the discontinuity of $M_+$ as the smoking gun of a first order phase transition. For a better perspective and comparison between $M_+$ and $M_-$, we have re-scaled them by their values at $\theta=0$ and combined their plots in Fig. \ref{fig:scaled_mass}. As $\theta$ increases the masses split and at a critical $\theta$, $M_+$ undergoes the discontinuity and changes sign. The red dots show the absolute value of $M_+$ above $\theta_c$. For this range, the two masses become the mirror image of one another. We notice that since the curves are re-scaled, they coincide for any value of $x>1$.

\begin{figure}
\includegraphics[width=8cm]{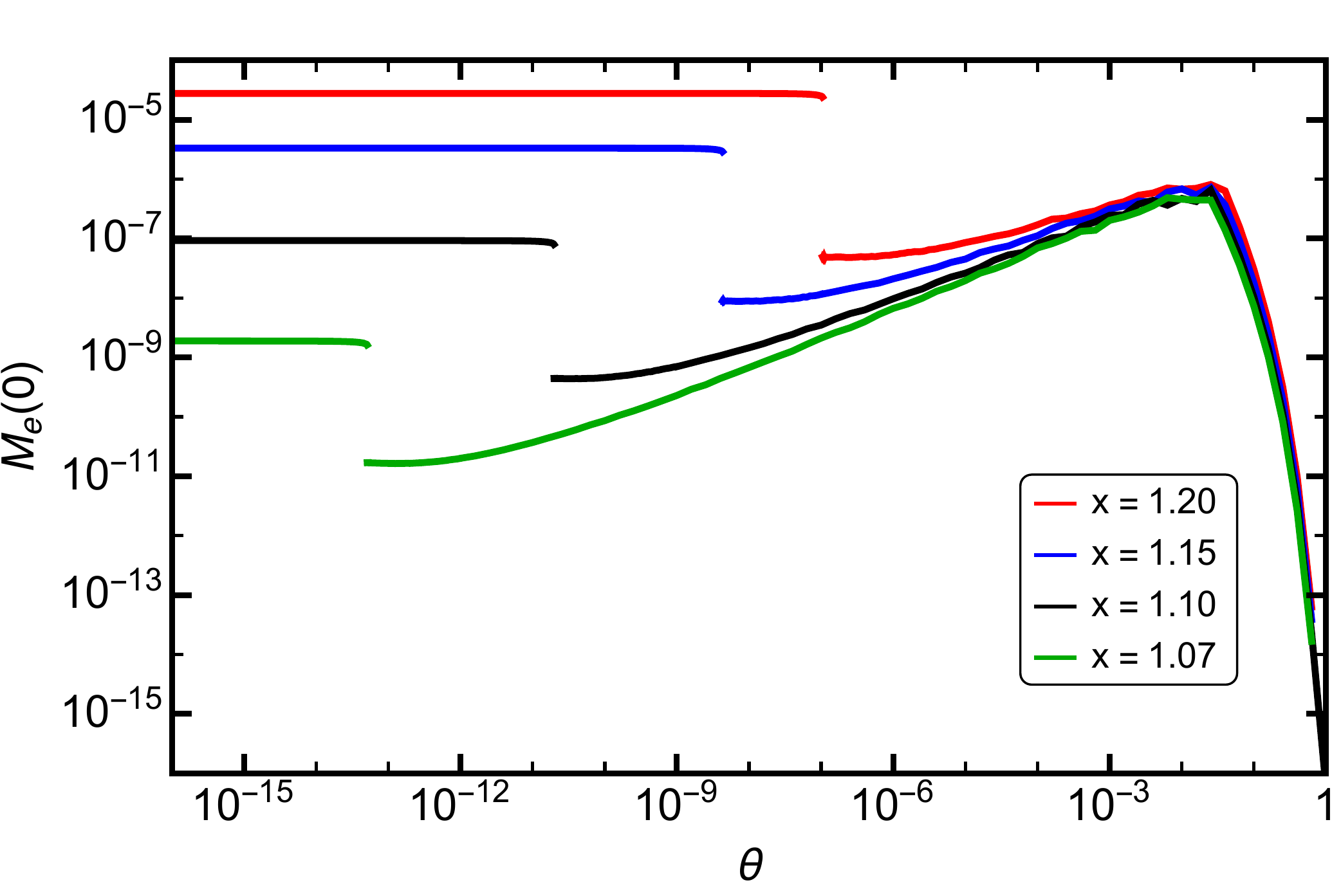}
\caption{Dynamical mass term, even under parity transformation, as a function of the CS parameter $\theta$}
\label{fig:m_even}
\end{figure}

\begin{figure}
\includegraphics[width=8cm]{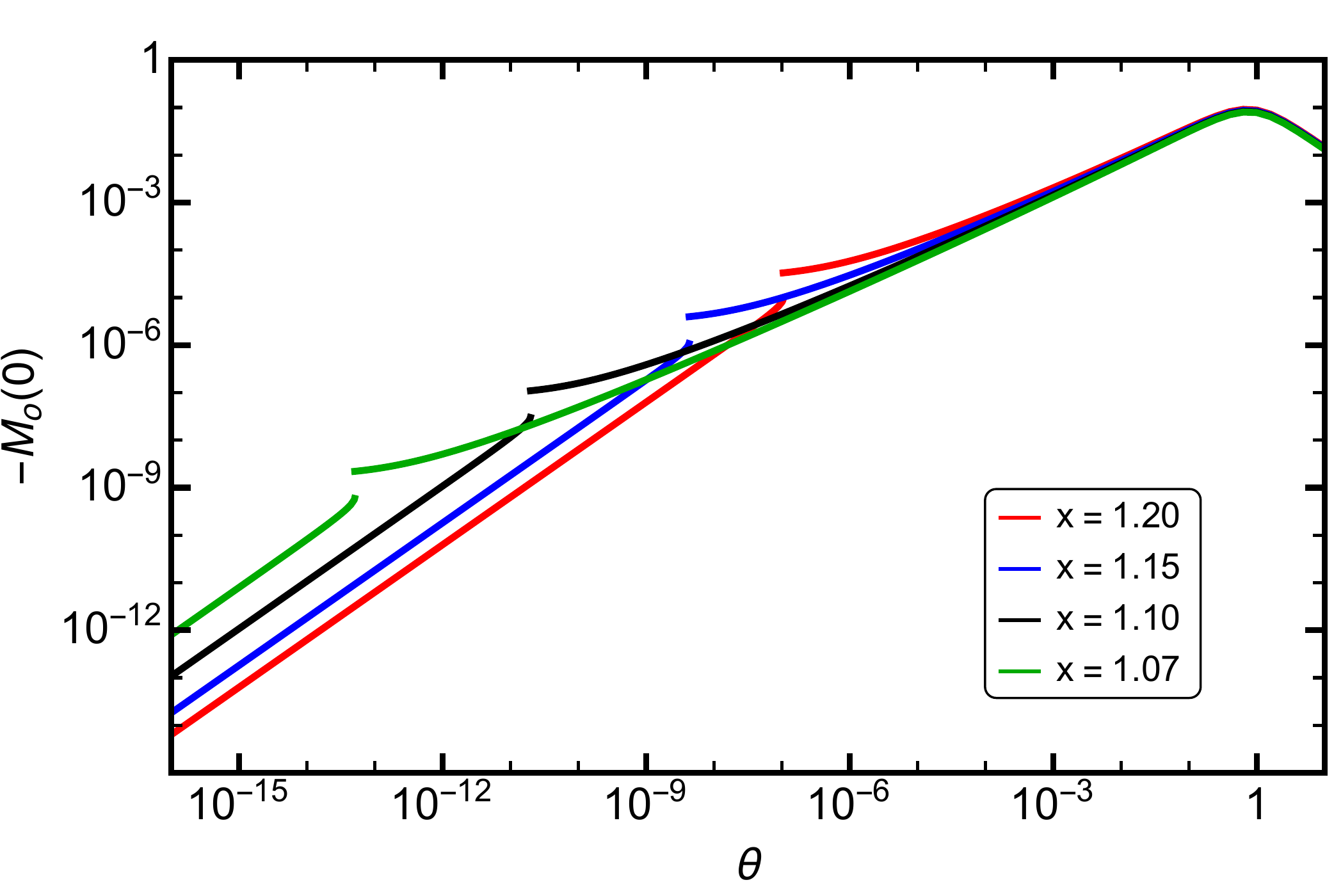}
\caption{Dynamical mass term, odd under parity transformation, as a function of the CS parameter $\theta$}
\label{fig:m_odd}
\end{figure}

The parity even and parity odd masses that appear in the Lagrangian (\ref{massive_Lagrangian}) can be obtained summing up or subtracting $M_\pm$, respectively. We can verify in Figs. \ref{fig:m_even} and \ref{fig:m_odd} that both, $m_e$ and $m_o$, are non-vanishing for a large range of $\theta$, clearly showing that CSB can occur within this model. For small $\theta$ the Dirac mass $m_e$, responsible for breaking chiral symmetry, presents a plateau, leaping at a critical $\theta$ to a value several orders of magnitude smaller. After that, the $m_e$ increases again and finally drops to zero. We interpret the discontinuity as an atempt to restore chiral symmetry, associated to the first piece in Eq.(\ref{mpm}). That would be the critical point if $f(\theta,0)$ was absent. However, as one moves towards the right of the discontinuity, the CS contribution $f(\theta,0)$ becomes dominant and $m_e$ starts to be significant again. For even larger $\theta$ the Dirac mass drops and the real critical $\theta$, for which chiral symmetry is restored, is around $\theta=1$. In turn, parity breaking is encoded in $m_o$. As the CS parameter becomes dominant, the Haldane mass increases, never restoring the symmetry.

 \section{Final remarks}
 \label{conclusions}
 In this paper we have investigated the chiral symmetry breaking and restoration in the framework of pseudo-QED (aka Reduced-QED) with a Chern-Simons (CS) term. This is a mixed dimensional theory applicable to systems where fermions are constrained to a plane while gauge fields are free to live in the bulk. The interaction of charge carriers in planar Dirac/Weyl materials and external electromagnetic fields in general must be described by this type of theories.
 
 We have calculated the radiative correction due to the CS, and suitable truncate non-perturbatively the SDE in order to explore the dynamical mass generation in the model. We worked in the quenched approximation and used rainbow-ladder truncation. In order to explore the CSB we searched for two types of masses, the Dirac mass $m_e\bar{\psi}\psi$ and the Haldane mass $m_o\bar{\psi}\tau\psi$. These bilinears are such that while the Dirac mass breaks chiral symmetry and preserves parity and time reversal, the Haldane mass leaves the chiral symmetry intact and violates these discrete symmetries. 
 
 We have shown that both masses appear for small values of $\theta$, provided the coupling considered is above the critical coupling $\alpha=\pi/8$. A competition between the pure P(R)QED and the effect of the CS term takes place, the latter giving contributions for both masses. For large enough $\theta$, the chiral symmetry is finally restored, while the Haldane mass remains finite and the breaking of $\mathcal{P}$ symmetry persists for all the range of $\theta$ considered here. 
 
 This result is specially relevant on what concerns chiral currents in planar materials. The chiral magnetic effect, initially proposed to take place in the quark gluon plasma formed in heavy ion collisions \cite{Kharzeev:2007jp,Fukushima:2008xe}, was observed in 3D Dirac materials \cite{Li:2014bha}. This was possible since the charge carriers in these kind of material have a relativistic-like behavior, allowing for a connection with high energy physics. The gapless nature of these quasi-particles makes their dispersion relation identical to the one of a massless free fermion. When an imbalance of chirality takes place, an anomalous electric current is generated in the direction of an applied external magnetic field. In the quark gluon plasma it is argued that the quantum anomaly responsible for this imbalance is originated from topological configurations of the gauge fields. In materials, the chiral anomaly is neatly generated applying simultaneously an electric and a magnetic field aligned.  
 
 It was considered by some of us that the chiral magnetic effect could happen in another class of materials, with potential advantages \cite{Mizher:2018dtf,Mizher:2013kza}: planar materials organized in honeycomb lattices. The standard way to model this type of structure is to work with two triangular sublattices and expand the energy around zero energy points, known as Dirac points. The two sublattices and two non-trivial Dirac points make a four-component spinor and produce an analog to the chirality. It was shown in \cite{Mizher:2018dtf,Mizher:2013kza} that as far as the fermion sector contains a Dirac and a Haldane mass, the chiral magnetic effect might occur in these semi-metals. 
 
The dynamically generated masses obtained in the present work correspond to the ones considered in \cite{Mizher:2018dtf,Mizher:2013kza}. The CS term have shown to be capable to give rise to Dirac and Haldane mass. Considerations excluding the possibility of the chiral magnetic effect in planar materials were taken within QED strictly in 3 space-time dimensions, while the suitable theory to model real materials immersed in external fields must be mixed dimensional. Since parallel electric and magnetic fields correspond to a CS, it might be possible that in-plane colinear fields applied in honeycomb lattice samples, like graphene, could generate the desired gap to work the chiral magnetic effect out.

We believe that the contribution of our results is two-fold: as a toy model to investigate anomaly effects in the chiral symetry breaking of QCD and as a model to describe gap generation in planar Dirac materials that happen via interaction with external electromagnetic fields.

 \section*{Acknowledgments}
A.~J.~Mizher receives partial support from FAPESP under fellowship number  2016/12705-7.
 
\appendix

\section{Obtaining the gap equation}
\label{appendixA}
For the chiral projectors, the following trace theorems follow:
\begin{eqnarray}
{\rm Tr}[\chi_\pm]&=&2\nonumber\\
{\rm Tr}[\gamma_\mu \chi_\pm]&=&0\nonumber\\
{\rm Tr}[\gamma_\mu \gamma_\nu\chi_\pm]&=&-2\delta_{\mu\nu}\nonumber\\
{\rm Tr}[\gamma_\mu \gamma_\nu \gamma_\rho \chi_\pm]&=& \mp 2 \epsilon_{\mu\nu\rho}\\
{\rm Tr}[\gamma_\mu\gamma_\nu\gamma_\rho\gamma_\sigma\chi_\pm]&=&2(\delta_{\mu\nu}\delta_{\rho\sigma}-\delta_{\mu\rho}\delta_{\nu\sigma}+\delta_{\mu\sigma}\delta_{\nu\rho}).\nonumber
\end{eqnarray}

The angular integral
			\begin{eqnarray}
			I_1&=&2\int_{0}^{\pi}\frac{d\theta\sin\theta}{\sqrt{k^2+p^2-2kp\cos\theta}}
			\end{eqnarray}
is written as
			\begin{eqnarray}
			I_1&=&\int_{-1}^{1}\frac{dx}{\sqrt{k^2+p^2-2kpx}}\nonumber\\
				&=&\frac{\theta(k-p)}{k}\int_{-1}^{1}\frac{dx}{\sqrt{1+\frac{p^2}{k^2}-2\frac{p}{k}x}}\\ \nonumber
				&&+\frac{\theta(p-k)}{p}\int_{-1}^{1}\frac{dx}{\sqrt{1+\frac{k^2}{p^2}-2\frac{k}{p}x}}.
			\end{eqnarray}
Then, it is simplified from the identity
			\begin{equation*}
				\int_{-1}^{1}\frac{dx}{\sqrt{1-a^2-2ax}}=2,
			\end{equation*}
valid $0<a<1$.
The angular integral
			\begin{eqnarray}
				I_2&=&\int_{0}^{\pi}\frac{d\theta\sin\theta k\cdot q}{k^2+p^2-2kp\cos\theta}\nonumber\\
				&=&\int_{0}^{\pi}\frac{d\theta\sin\theta(k^2-kp\cos\theta)}{k^2+p^2-2kp\cos\theta}
			\end{eqnarray}
can be conveniently written as
\begin{eqnarray}
				I_2
				&=&k^2\Big[\frac{\theta(k-p)}{k^2}\int_{-1}^{1}\frac{dx}{1+\frac{p^2}{k^2}-2\frac{p}{k}x}\nonumber\\
				&&+\frac{\theta(p-k)}{p^2}\int_{-1}^{1}\frac{dx}{1+\frac{k^2}{p^2}-2\frac{k}{p}x}\Big]\nonumber\\
				&-&kp\Big[\frac{\theta(k-p)}{k^2}\int_{-1}^{1}\frac{dx x}{1+\frac{p^2}{k^2}-2\frac{p}{k}x}\nonumber\\
				&&+\frac{\theta(p-k)}{p^2}\int_{-1}^{1}\frac{dx x}{1+\frac{k^2}{p^2}-2\frac{k}{p}x}\Big].
			\end{eqnarray}
Then, the final form of the gap equation is obtained from 
\begin{equation}
\int_{-1}^{1}\frac{dx}{1+a^2-2ax}\approx 2 
\end{equation}
and
\begin{equation}
\int_{-1}^{1}\frac{dx x}{1+a^2-2ax}\approx\frac{4a}{3} ,
\end{equation}			 
where the asymptotic behavior is reached as $ (a\rightarrow 0).$

 \end{document}